\documentclass[%
superscriptaddress,
frontmatterverbose, 
nofootinbib,
 amsmath,amssymb,
 aps,
prd,
twocolumn,
notitlepage
]{revtex4-1}

\usepackage{graphicx}
\usepackage{dcolumn}
\usepackage{bm}
\usepackage{units}

\begin{document}

\title{Quantum simulation of traversable wormhole spacetimes in a Bose-Einstein condensate}
\author{Jes\'us Mateos}
\affiliation{Facultad CC. F\'isicas, Universidad Complutense of Madrid,Plaza Ciencias, 1 Ciudad Universitaria 28040 Madrid (Spain)}\email{jesmateo@ucm.es}
 \author{Carlos Sab\'in}
 \affiliation{Instituto de F\'isica Fundamental, CSIC, Serrano, 113-bis, 28006 Madrid (Spain)}
\begin{abstract}
In this work we propose a recipe for the quantum simulation of traversable wormhole spacetimes in a Bose-Einstein condensate, both in $1+1 D$ and $3+1 D$. While in the former case it is enough to modulate the speed of sound along the condensate, in the latter case we need to choose particular coordinates, namely generalized Gullstrand-Painlev\'e coordinates. For weakly interacting condensates, in both cases we present the spatial dependence of the external magnetic field which is needed for the simulation, and we analyze under which conditions the simulation is possible with the experimental state-of-the-art.
\end{abstract}

\maketitle

\section{Introduction}

Quantum simulators enable to study properties of quantum systems which are otherwise out of experimental reach, by mimicking them with more experimentally amenable quantum systems. In this sense, among other possible approaches, they may be thought of as a window to the analysis of physics lying at the edges of the theory \cite{zitterbewegung,Sabin2,gws2}, and even beyond \cite{majoranas1,majoranas2,tachions1,tachions2}.

Traversable wormholes are interesting cosmological objects which appear in certain solutions of general relativity equations. In principle, they connect distant regions of the universe, or even regions of different universes. Due to this behavior as a spacetime shortcut, they are the focus of great theoretical interest, and they are used as a pedagogical tool in general relativity \cite{Thorne, Thorne2, Libro1, Libro2}. In particular, they are proposed and studied like a way for interstellar travels \cite{Thorne}. Recently, they have generated renewed attention as possible ``black hole mimickers" \cite{gravastar,konoplya}. 

However, today their existence has not been demonstrated in a direct or indirect observational way. Moreover, it seems that they do not appear naturally in our universe, and there are theoretical reasons to expect that they must be forbidden \cite{Hawking}. Indeed, in order for these objects to be stable, they must be made of exotic material which violate the weak energy condition \cite{Thorne, Thorne2} and these spacetimes may contain closed timelike curves (CTCs), which might imply a violation of the principle of causality \cite{Thorne2}. In this sense, a quantum simulator can be a useful tool in the analysis of traversable wormholes. Indeed, recently one of us has proposed a quantum simulator of a traversable wormhole spacetime by means of a dc-SQUID array \cite{Sabin1}. This proposal was restricted to $1+1 D$ and to a particular type of wormhole, namely the Ellis wormhole. 

Bose-Einstein condensates (BEC) haven been used widely in quantum simulation of cosmological objects, namely black holes \cite{Garay1, Garay2} or gravitational waves \cite{Sabin2}.

In this work, we show how to simulate a variety of traversable wormhole spacetimes in a BEC, both in $1+1D$ --where we go beyond the Ellis wormhole case-- and in the more realistic $3+1D$ case. In the former case, it is enough to modulate the speed of sound along the condensate by means of a Feshbach resonance. In the latter, we introduce particular coordinates to achieve the simulation, namely generalized Gullstrand-Painlev\'e (GP) coordinates. We will analyze in detail the prospects for an experimental implementation with current technology.

\begin{figure}[t]
   \centering
\includegraphics[scale=1]{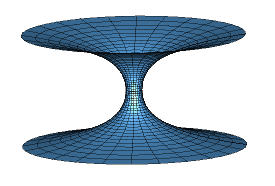}
 \caption{\footnotesize Embedding diagram for an Ellis wormhole (shape function given in \eqref{ec:b} with $q=-1$) with radius of the throat $b_0=3$. The upper part corresponds to $l>0$, while the lower part corresponds to $l<0$. These asymptotically flat regions are connected by the wormhole's throat at $r=b_0$ ($l=0$). This embedding diagram is derived following the standard techniques given in \cite{Thorne}.}
     \label{fig:elliswh}
\end{figure}

\section{Traversable wormhole spacetimes}

We start from the line element of a traversable wormhole spacetime, which is given by \cite{Thorne, Sabin1}
\begin{align}
\label{ec:dsw}
ds^2=-c^2e^{2\phi(r)}dt^2+\frac1{1-\frac{b(r)}{r}}dr^2+r^2(d\theta^2+\sin^2\theta d\phi^2),
\end{align}
where $\phi(r)$ is the redshift function and $b(r)$ is the shape function, and both functions depend on the radius $r$ only. The features of the wormhole are fully determined by these two functions. In particular, $\phi$ and $b$ can be adjusted in order to allow a travel trough the wormhole. In the case of the shape function, there exists a value $r=b_0$ such that $b(r=b_0)=r=b_0$, which determines the position of the wormhole's throat. It defines the proper radial distance $l$ to the throat as $l=\pm\int^r_{b_0}dr'\left(1-b(r')/r'\right)^{-1/2}$, which defines two different regions into the same universe for $l>0$, corresponding to $r$ going from $\infty$ to $b_0$, and $l<0$, corresponding to $r$ going from $b_0$ to $\infty$. Therefore, when $r\to\infty$, we have two asymptotically flat regions corresponding to $l\to\pm\infty$, which are connected through the wormhole throat at $l=0$, i.e. at $r=b_0$. The embedding diagram of Fig. \ref{fig:elliswh} shows a pictorial way to understand these concepts.

For simplicity, we will study a massless wormhole, i.e. $\phi(r)=0$, and therefore the expression \eqref{ec:dsw} reduces to
\begin{align}
\label{ec:dsw3d}
ds^2=-c^2dt^2+\dfrac1{1-\frac{b(r)}{r}}dr^2+r^2(d\theta^2+\sin^2\theta d\phi^2),
\end{align}
and now the wormhole is characterized by $b(r)$ only.

On the other hand, the effective metric of a BEC in $3+1$ dimensions is a curved metric, which is given as follows \cite{Sabin2, Fagno, Visser1}
\begin{align}
\label{ec:dsbec3}
G_{\mu\nu}=\frac{\rho c}{c_s}\left[g_{\mu\nu}+\left(1-\frac{c_s^2}{c^2}\right)\frac{v_\mu v_\nu}{c^2}\right],
\end{align}
where $g_{\mu\nu}$ is the real spacetime metric in which the condensate is, that it can be curved in general. In our case we take $g_{\mu\nu}=\eta_{\mu\nu}$, where $\eta_{\mu\nu}$ is the flat Minkowski metric (note that in this work we choose the ``mostly plus" convention for the signature of $\eta_{\mu\nu}$). Furthermore, $\rho$ is the density of the BEC, $c_s$ is  the phonon propagation speed in the condensate, and $v_\mu$ is the velocity flow 4-vector, which is associated to the 4-divergence of the BEC's phase \cite{Fagno}.

In order to simulate a traversable wormhole in the BEC, the aim is to relate the effective metric \eqref{ec:dsbec3} with \eqref{ec:dsw3d}.

\section{$\boldsymbol{1+1D}$ case}

For simplicity, we work first with the $1+1$ dimensional case.  Here, we can leverage the conformal invariance of the Klein-Gordon equation in $1+1 D$.  The line element (\ref{ec:dsw}) becomes
\begin{equation}
ds^2=-c^2dt^2+\frac1{1-\frac{b(r)}{r}}dr^2,
\end{equation}
that is conformal to
\begin{align}
\label{ec:dsw1d}
ds^2=-c^2\left[1-\frac{b(r)}{r}\right]dt^2+dr^2=-c^2(r)dt^2+dr^2,
\end{align}
i.e. we found a metric with an effective speed of light which depends on the radial distance as follows
\begin{align}
\label{ec:ceffw}
c^2(r)=c^2\left[1-\frac{b(r)}{r}\right].
\end{align}
Notice that we are not interested in this spacetime \textit{per se}, but only as a $1+1 D$ section of a $3+1 D$ spacetime. Of course, there is no throat in one spatial dimension, but only a one-dimensional section of it, namely a point.

Meanwhile, the metric (\ref{ec:dsbec3}) of a condensate in $1+1$ dimensions, and considering that the velocity flow is just $v^\mu=(c, 0)$, reduces to
\begin{eqnarray}G_{\mu\nu}&=&\frac{\rho c}{c_s}\left[\begin{pmatrix}-1 & 0\\0 & 1\end{pmatrix}+\left(1-\frac{c_s^2}{c^2}\right)\begin{pmatrix}1 & 0\\0 & 0\end{pmatrix}\right]\nonumber\\ &=&\frac{\rho c}{c_s}\begin{pmatrix}-(c_s/c)^2 & 0\\0 & 1\end{pmatrix},\end{eqnarray}
and therefore the line element is conformal to
\begin{align}
\label{ec:dsbec1d}
ds^2=-\frac{c_s^2}{c^2}c^2dt+dr^2=-c_s^2dt^2+dr^2.
\end{align}

To achieve that the metric (\ref{ec:dsbec1d}) simulates the target metric (\ref{ec:dsw1d}), the task is to modulate the speed of sound in the BEC, that we suppose in general $c_s=c_{s0}f(r)$, with $c_{s0}$ constant, as the expression (\ref{ec:ceffw}) of the effective speed of light at the wormhole. It is important to remark that we do not simulate a real spacetime, but an acoustic spacetime in which $c_s$ plays the role of $c$.

In a weakly interacting condensate, the speed of sound depends on the coupling strength $g$, which in turn depends on the scattering length $a$ of the BEC as follows \cite{Sabin2, Garay2}
\begin{align}
\left.\begin{array}{ll}
c_s=\sqrt{\dfrac{\rho g}{m}}\\ \\
g=\dfrac{4\pi\hbar^2a}{m}
\end{array}
\right\}\quad \Rightarrow\;c_s=c_s(a)=\frac{\hbar}{m}\sqrt{4\pi\rho a},
\label{ec:csa}
\end{align}
where $m$ is the atomic mass of the BEC and $\rho$ is its density. 
\begin{figure*}[t]
     \label{fig:aBplots}
\includegraphics[width=\textwidth]{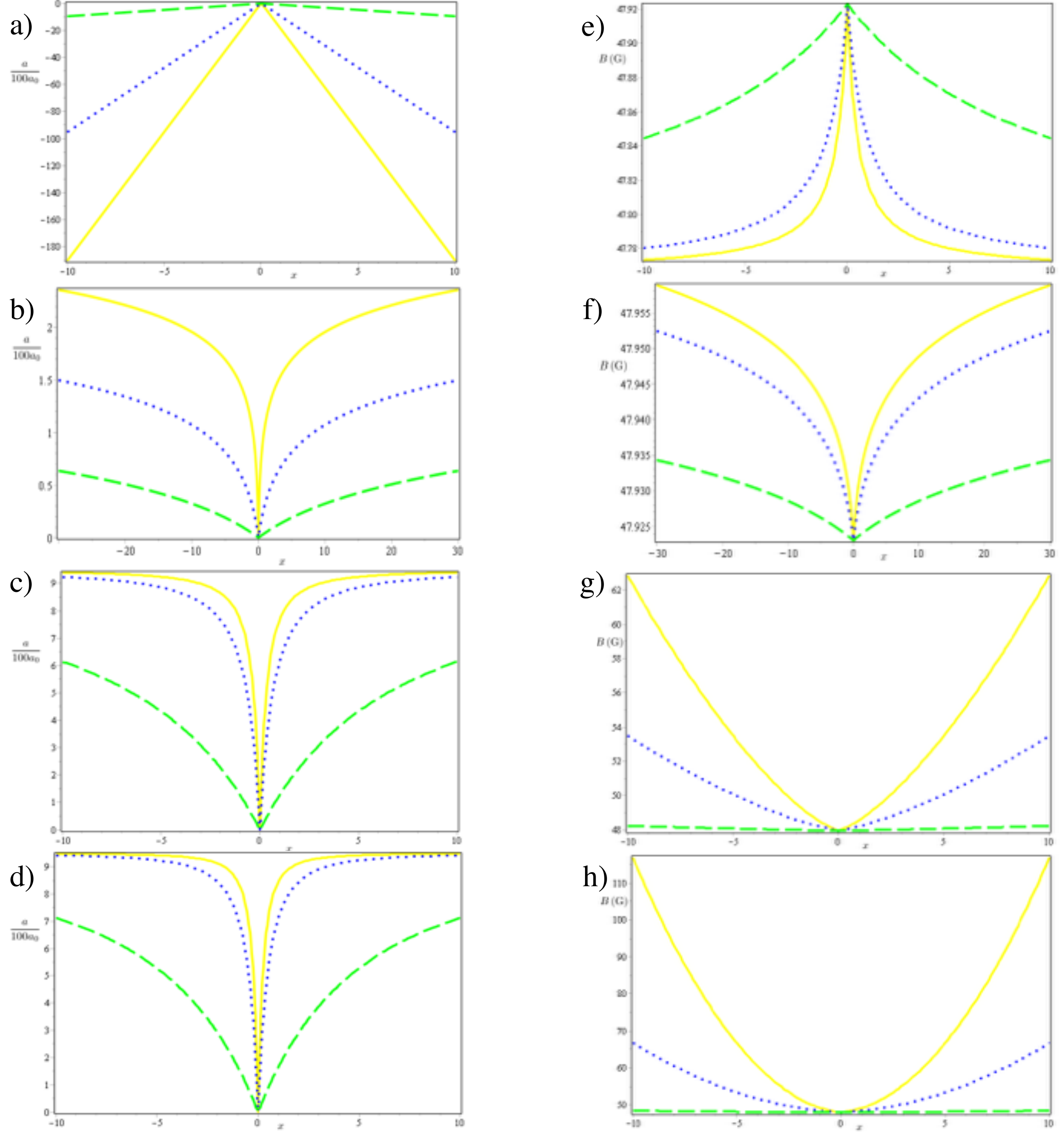}
 \caption{\footnotesize Scattering length $a$ (a)-(d) and magnetic field $B$ (e)-(h) vs. $x$, for several values of $q$ and $b_0$. The value of $q$ is  $q=2$ (a) and (e),$q= 0.95$ (b) and (f), $q=-0.5$ (c) and (g),$q= -1$ (d) and (h). In all the plots: $b_0=0.5$ (yellow, solid), $b_0=1$ (blue, dotted), $b_0=10$ (green, dashed).}
     \label{fig:aBplots}
\end{figure*}

Near the Feshbach resonance, $a$ depends on the external magnetic field $B$ \cite{Sabin2, Exp}
\begin{align}
\label{ec:aB}
a=a_{bg}\left(1-\frac{\omega}{B-B_0}\right),
\end{align}
where $a_{bg}$ is the background scattering length, $\omega$ is the width of the resonance and $B_0$ is the value of $B$ at which the resonance takes place. Replacing (\ref{ec:aB}) in (\ref{ec:csa}) we have
\begin{align}
\label{ec:cB}
c_s=c_s(B)=c_{s0}\sqrt{{1-\frac{\omega}{B-B_0}}},
\end{align}
where $c_{s0}=(\hbar/m)\sqrt{4\pi \rho a_{bg}}$ is constant.

Now, we have to found the explicit dependence of $B$ on the distance $r$, in order to relate (\ref{ec:cB}) with  (\ref{ec:ceffw}). If we equate these expressions, we have
\begin{align}
\frac{\omega}{B-B_0}=&\frac{b(r)}{r}\Rightarrow \omega=\frac{b(r)}{r}(B-B_0)\nonumber\\
\Rightarrow&B(r)=\frac{r}{b(r)}\omega+B_0,
\label{ec:Br1}
\end{align}
so the external field depends on the explicit form of the shape function $b(r)$.

Then we choose as shape function of the wormhole \cite{taylor}
\begin{align}
\label{ec:b}
b(r)=b_0^{1-q}r^q,
\end{align}
where $b_0$ is the wormhole throat radius. Replacing this in (\ref{ec:Br1}), $B$ becomes
\begin{align}
\label{ec:Br2}
B(r)=\frac{r^{1-q}}{b_0^{1-q}}\omega+B_0,
\end{align}
where we have the freedom that offers us the radius $b_0$ and the parameter $q$. Note that $q=-1$ corresponds to the Ellis wormhole case.

In order to choose the best value of $q$, we compare the curve
\begin{align}
\label{ec:ar1}
\frac{a(r)}{a_{bg}}=1-\frac{\omega}{B(r)-B_0}=1-\frac{b_0^{1-q}}{r^{1-q}}
\end{align}
which is the result of replacing \eqref{ec:aB} in \eqref{ec:Br2}, with state-of-the-art experiments for the spatial variation of the scattering length \cite{Exp}.

Now, taking the following experimental values of a cesium BEC from \cite{Exp} $\omega=\unit[157]{mG}$, $B_0=\unit[47.766]{G}$ and $a_{bg}\simeq 950a_0$, where $a_0$ is the Bohr radius, we can plot the curves $a(r)/(100a_0)$ and $B(r)$ from (\ref{ec:ar1}) and (\ref{ec:Br2}), respectively, for several values of $b_0$ and $q$. For the sake of convenience, we do a final step before the plots, and this is to define a new spatial coordinate such that \cite{Sabin1}
\begin{align}
\label{ec:x}
|x|=r-b_0,\quad x\in(-\infty, \infty).
\end{align}
Clearly, $x=0$ at the wormhole's throat $r=b_0$, and acquires different sign at both sides of the throat.

With all these considerations, we obtain the plots which are given in the Fig. \ref{fig:aBplots}. We observe that the plots of $a(r)/100a_0$ [\ref{fig:aBplots}(a)-(d)] for the values $q<1$ are similar to the experimental Fig. 4  in \cite{Exp}. Moreover, for the interval $0<q<1$ we obtain a maximum value for $a/100a_0$ which is very close to the experimental one [see in particular \ref{fig:aBplots}(b) and (f) for the value $q=0.95$]. Of course, in our case the curve looks less smooth than the experimental one as we get close to the throat.
Besides, for $0<q<1$, we found more similarities with Fig. 4 in \cite{Exp} for small values of $b_0$. This is due to the fact that the ratio $b_0/r$ is dimensionless, and hence the size of the condensate determines the size of the throat. So, green lines represent wormhole's scales bigger than its corresponding BEC's scale in the plot.
However, a good figure of merit can be the variation of scattering length per unit length of the BEC. By inspection of the 0 to 15 $\mu$m region in Fig 4c) of \cite{Exp}, in which the largest variation occurs, we find that they can vary the parameter $a/(100a0)$ an amount of 0.067 per micron, approximately. Referring to our case, differentiating \eqref{ec:ar1}, with the change \eqref{ec:x}, and evaluating this expression for the values of $q$ and $b_0$ which provides the most similar plot to the experimental one, i.e. $q=0.95$, $b_0=1$, and taking $x=10$ --the range of $x$ for which we have the greatest increase of this parameter-- we obtain a variation of $0.038$ per micron. So, we have that our requirements are totally compatible with the capability shown in \cite{Exp}.

\begin{figure}[t]
   \centering
\includegraphics[scale=0.5]{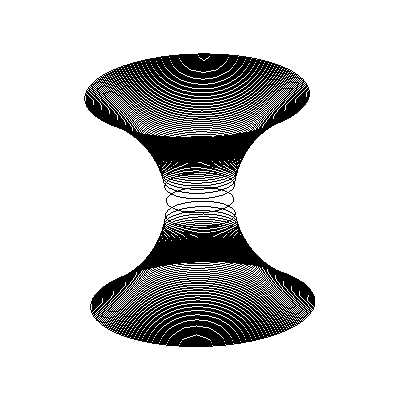}
 \caption{\footnotesize Numerical embedding diagram for a wormhole spacetime with shape function given in \eqref{ec:b} with $q=0.5$, and with radius of the throat $b_0=3$. This embedding diagram is derived following the standard techniques given in \cite{Thorne}. This value of $q$ is within the interval $0<q<1$ for which we achieve the simulation in the one-dimensional case.}
     \label{fig:whq05}
\end{figure}

Therefore the family of wormhole spacetimes corresponding to the range $0<q<1$ could be simulated in the lab with the technology of \cite{Exp}. We want to remark that the value $q=-1$ corresponds to the Ellis wormhole \cite{Ellis}, which is the most frequent in the literature \cite{Thorne2, Sabin1, whgeodesics1, whgeodesics2}, and whose simulation has already been proposed in the one dimensional case in \cite{Sabin1}. Therefore, it is important to note that we propose a simulation of a different one-dimensional kind of wormhole. An example of embedding diagram is given in Fig. \ref{fig:whq05}.

\section{$\boldsymbol{3+1D}$ case}\label{sec3}

This case is more sophisticate than the former. Now a modulation of $c_s$ is not enough, because (\ref{ec:dsw3d}) and (\ref{ec:dsbec3}) are more involved than in the one-dimensional case. In particular, we are no longer able to exploit the conformal invariance of the Klein-Gordon equation.

We start by writing the real background metric in (\ref{ec:dsbec3}) $g_{\mu\nu}=\eta_{\mu\nu}$ in spherical coordinates, in analogy with (\ref{ec:dsw3d}). Now, for the velocity flow of the BEC  without lack of generality we take $v^\mu=(c, v^r, 0, 0)$. 
With all these considerations, the metric (\ref{ec:dsw3d}) becomes
\begin{eqnarray}
\label{ec:Gnodiag}
&&G_{\mu\nu}=\frac{\rho c}{c_s}\nonumber\\&&\begin{pmatrix}
-\frac{c_s^2}{c^2} & -[1-\frac{c_s^2}{c^2}]\frac{v^r}{c} & 0 & 0\\
 -[1-\frac{c_s^2}{c^2}]\frac{v^r}{c}&  1+[1-\frac{c_s^2}{c^2}]\frac{(v^r)^2}{c^2} & 0& 0\\
0 & 0 & r^2 & 0\\
0 & 0 & 0 & r^2\sin^2\theta\\
\end{pmatrix}
\end{eqnarray}
where the covariant components of the contravariant 4-vector $v^\mu$ are $v_\mu=(-c, v^r, 0, 0)$, which explains the minus sign in the nondiagonal elements.

Therefore, now we have a nondiagonal metric for the BEC, through which we want to simulate the diagonal wormhole metric \eqref{ec:dsw3d}. 
In order to do it, we introduce a change of coordinates for the wormhole spacetime, which provides a nondiagonal metric.

We choose coordinates based on the Gullstrand-Painlevé (GP) ones \cite{Painleve, Gullstrand}, which were originally introduced for the Schwarzschild black hole. For these coordinates, the new time coordinate follows the proper time of a free-falling observer who starts from far away at zero velocity.

We will construct GP-like coordinates for the wormhole spacetime based on the generalized GP coordinates given in \cite{GP}, for more general observers with velocity $v_\infty\neq 0$ in the infinity, which falls towards the hole following geodesics.

So we start from the timelike geodesics of the metric \eqref{ec:dsw3d}, for the Ellis wormhole case in which the shape function is $b(r)=b_0^2/r$. These geodesics are given, with $c=1$, in their first-order form, by \cite{whgeodesics1,whgeodesics2}
\begin{align}
&\dot t=E,\\
&1+\frac{\dot r^2}{1-\frac{b_0^2}{r^2}}+\frac{L^2}{r^2}=E^2,
\end{align}
where $E$ and $L$ are the energy and the angular momentum per unit mass of the observer, respectively. This energy satisfies $E\geq 1$ and its related to $v_\infty$ by
\begin{align}
\label{ec:Evinfty}
E=\frac1{\sqrt{1-v^ 2_\infty}}.
\end{align}

For simplicity, we consider radial geodesics, hence $L=0$. So we have
\begin{align}
\label{ec:geodesict}
&\dot t=E,\\
\label{ec:geodesicr}
&\dot r=\pm\sqrt{\left(1-\frac{b_0^2}{r^2}\right)\left(E^2-1\right)},
\end{align}
where we choose the minus sign, which corresponds to ingoing geodesics.

Therefore, the 4-velocity of the observer is $u^\mu=(\dot t,\dot r,0,0)$, and its covariant counterpart is given as follows
\begin{align}
u_\mu=\left(-E,-\sqrt{\frac{E^2-1}{1-\frac{b_0^2}{r^2}}},0,0\right),
\end{align}
so we can identify it with the gradient of a new temporal GP-like coordinate $t_r$, i.e. $u_\mu=-\partial_\mu t_r$, which is given by
\begin{align}
\label{ec:tr}
t_r=Et+{\int} dr\sqrt{\frac{E^2-1}{1-\frac{b_0^2}{r^2}}}.
\end{align}
From this, we have
\begin{align}
\nonumber
dt&=\frac1{E}dt_r-\frac1{E}\sqrt{\frac{E^2-1}{1-\frac{b_0^2}{r^2}}}dr,\\
\label{ec:dtr2}
dt^2&=\frac1{E^2}dt_r^2+\frac1{E^2}\left(\frac{E^2-1}{1-\frac{b_0^2}{r^2}}\right)dr^2-\frac2{E^2}
\sqrt{\frac{E^2-1}{1-\frac{b_0^2}{r^2}}}dt_rdr,
\end{align}
and replacing \eqref{ec:dtr2} in \eqref{ec:dsw3d}, the wormhole line element in these GP-like coordinates is
\begin{equation}
ds^2=-\frac1{E^2}dt_r^2+\frac2{E^2}
\sqrt{\frac{E^2-1}{1-\frac{b_0^2}{r^2}}}dt_rdr+
\frac{1}{E^2\left(1-\frac{b_0^2}{r^2}\right)}dr^2+r^2d\Omega^2.
\end{equation}
In order to recover units, we must take into account first that the energy per unit mass is given by
\begin{align}
\label{ec:Eunits}
E=\gamma c^2=\frac{c^2}{\sqrt{1-\left(\frac{v_\infty}{c}\right)^2}},\end{align}
which has units of $c^2$. The line element must have units of square length, and the metric elements must be dimensionless, so we have
\begin{eqnarray}
\label{ec:dsGPE}
ds^ 2&=&-\frac{c^4}{E^2}c^2dt_r^2+2\frac{c^4}{E^2}
\sqrt{\frac{\frac{E^2}{c^4}-1}{1-\frac{b_0^2}{r^2}}}cdt_rdr\nonumber\\&+&
\frac{1}{\frac{E^2}{c^4}\left(1-\frac{b_0^2}{r^2}\right)}
dr^2+r^2d\Omega^2,
\end{eqnarray}
which can be expressed in terms of the Lorentz factor $\gamma$, through the expression \eqref{ec:Eunits}, as follows
\begin{eqnarray}
\label{ec:dsGP}
ds^ 2&=&-\frac{c^2}{\gamma^2}dt_r^2+\frac2{\gamma^2}
\sqrt{\frac{\gamma^2-1}{1-\frac{b_0^2}{r^2}}}cdt_rdr\nonumber\\&+&
\frac{1}{\gamma^2\left(1-\frac{b_0^2}{r^2}\right)}
dr^2+r^2d\Omega^2.
\end{eqnarray}
Note that in the case of a free-falling observer, i.e. $v_\infty=0$ and hence $\gamma=1$, we recover the original line element of the wormhole \eqref{ec:dsw3d}, so this case is not interesting for us.

We can see that we have, like in the 1-dimensional case [see \eqref{ec:ceffw}], an effective speed of light 
\begin{align}
c_{\text{eff}}=\gamma^{-1}c,
\end{align}
so we can rewrite \eqref{ec:dsGP} in terms of $c_{\text{eff}}$ as follows
\begin{eqnarray}
\label{ec:dsGP2}
ds^ 2&=&-c^2_{\text{eff}}dt_r^2+\frac2{\gamma}
\sqrt{\frac{\gamma^2-1}{1-\frac{b_0^2}{r^2}}}c_{\text{eff}}dt_rdr\nonumber\\&+&
\frac{1}{\gamma^2\left(1-\frac{b_0^2}{r^2}\right)}
dr^2+r^2d\Omega^2.
\end{eqnarray}

In fact we can reproduce the wormhole spacetime in the condensate, not in a real spacetime, therefore we have to write an acoustic wormhole spacetime, in our GP-like coordinates, i.e. we have to replace in \eqref{ec:dsGP2} $c_{\text{eff}}$ and $\gamma$ by
\begin{align}
\label{ec:cseff}
c_{s_\text{eff}}&=\gamma_s^{-1}c_{s_0},\\
\gamma_s&=\frac1{\sqrt{{1-\frac{v^2_\infty}{c^2_{s_0}}}}},
\end{align}
(note that we have performed the same step, in an implicit way, at the one-dimensional case) so we find the following acoustic wormhole spacetime line element
\begin{eqnarray}
\label{ec:dsGPacoustic}
ds^ 2&=&-c^2_{s_\text{eff}}dt_r^2+\frac2{\gamma_s}
\sqrt{\frac{\gamma_s^2-1}{1-\frac{b_0^2}{r^2}}}c_{s_\text{eff}}dt_rdr\nonumber\\&+&
\frac{1}{\gamma_s^2\left(1-\frac{b_0^2}{r^2}\right)}
dr^2+r^2d\Omega^2.
\end{eqnarray}

Now, we compare this line element to the line element of the BEC, which is given, from \eqref{ec:Gnodiag}, as follows
\begin{eqnarray}
\label{ec:dsBEC3d}
ds^ 2&=&-c_s^ 2dt^ 2-2\Bigg[1-\left(\frac{c_s}{c}\right)^ 2\Bigg]\frac{v^r}{c}cdtdr\nonumber\\&+&
\Bigg\{1+\bigg[1-\left(\frac{c_s}{c}\right)^ 2\bigg]
\left(\frac{v^r}{c}\right)^2\Bigg\}dr^2+r^2d\Omega^2,
\end{eqnarray}
and, in order to achieve an acoustic wormhole, we have to rewrite the nondiagonal term like $g_{rt}c_sdtdr$
\begin{align}
\Bigg[1-\left(\frac{c_s}{c}\right)^ 2\Bigg]\frac{v^r}{c}cdtdr=
-v^r\Bigg[\frac1{c_s}-\frac{c_s}{c^2}\Bigg]c_sdtdr.
\end{align}
Equating component to component of both metrics, we obtain
\begin{itemize}
\item from $g_{tt}$
\begin{align}
\label{ec:csgamma}
c_s=c_{s_\text{eff}}=\frac{c_{s_0}}{\gamma_s}=c_{s_0}\sqrt{1-\frac{v^2_\infty}{c_{s_0}^2}},
\end{align}
\item from $g_{tr}$
\begin{align}
\frac2{\gamma_s}\sqrt{\frac{\gamma_s^2-1}{1-\frac{b_0^2}{r^2}}}c_{s_\text{eff}}dt_rdr=2v^r\Bigg[\frac1{c_{s_\text{eff}}}-\frac{c_{s_\text{eff}}}{c^2}\Bigg]c_{s_\text{eff}}dtdr,
\end{align}
\item from $g_{rr}$
\begin{align}
\frac{1}{\gamma_s^2\left(1-\frac{b_0^2}{r^2}\right)}
dr^2=
\Bigg\{1+\bigg[1-\left(\frac{c_s}{c}\right)^ 2\bigg]
\left(\frac{v^r}{c}\right)^2\Bigg\}dr^2,
\end{align}
\end{itemize}
so we have the following system of equations

\begin{subequations}
\begin{align}
\label{ec:system1}
&\frac1{\gamma_s}\sqrt{\frac{\gamma_s^2-1}{1-\frac{b_0^2}{r^2}}}=v^r\Bigg(\frac{\gamma_s}{c_{s_0}}-\frac{c_{s_0}}{\gamma_s c^2}\Bigg)\\\label{ec:system2}
&\frac{1}{\gamma_s^2\left(1-\frac{b_0^2}{r^2}\right)}
=
1+\bigg[1-\left(\frac{c_{s_0}}{c\gamma_s}\right)^ 2\bigg]
\left(\frac{v^r}{c}\right)^2
\end{align}
\end{subequations}

Now, in order to solve this system, we take into account that $$\frac{v^r}{c}, \frac{c_{s_0}}{c}\ll 1,$$
so we can approximate the previous system \eqref{ec:system1},\eqref{ec:system2} to zero order in $v^r/c$ and $c_{s_0}/c$. Hence we get
\begin{itemize}
\item for the left-hand side of \eqref{ec:system1}
\begin{align*}
\frac1{\gamma_s}\sqrt{\frac{\gamma_s^2-1}{1-\frac{b_0^2}{r^2}}}
=\frac{v_\infty}{c_{s_0}}\sqrt{\frac1{1-\frac{b_0^2}{r^2}}},
\end{align*}
\item for the right-hand side of \eqref{ec:system1}
\begin{align*}
\frac{v^r\gamma_s}{c_{s_0}}-\frac{v^rc_{s_0}}{\gamma_sc^2}\simeq
\frac{v^r\gamma_s}{c_{s_0}}=\frac{v^r}{c_{s_0}}\sqrt{\frac1{1-\frac{b_0^2}{r^2}}},
\end{align*}
\item for the left-hand side of \eqref{ec:system2}
\begin{align*}
\frac{1}{\gamma_s^2\left(1-\frac{b_0^2}{r^2}\right)}=\frac{1-\frac{v^2_\infty}{c_{s_0}^2}}{1-\frac{b_0^2}{r^2}},
\end{align*}
\item for the right-hand side of \eqref{ec:system2}
\begin{eqnarray*}
1&+&\left(1-\frac{c_{s_0}^2}{\gamma_s^2c^2}\right)\left(\frac{v^r}{c}\right)^2=
1+\left(\frac{v^r}{c}\right)^2\left(1-\frac{b_0^2}{r^2}\right)\left(\frac{c_{s_0}v^r}{c^2}\right)^2\nonumber\\\simeq &1&
\end{eqnarray*}
\end{itemize}
Therefore, we finally find

\begin{subequations}
\begin{eqnarray}
\label{ec:system1.1}
&\dfrac{v_\infty}{c_{s_0}}\sqrt{\dfrac1{1-\frac{b_0^2}{r^2}}}=\dfrac{v^r}{c_{s_0}}\sqrt{\dfrac1{1-\frac{b_0^2}{r^2}}}\\\label{ec:system2.1}
&\hspace{-2.5cm}\dfrac{1-\frac{v^2_\infty}{c_{s_0}^2}}{1-\frac{b_0^2}{r^2}}=1
\end{eqnarray}
\end{subequations}
\begin{figure*}[t]
\includegraphics[scale=0.4]{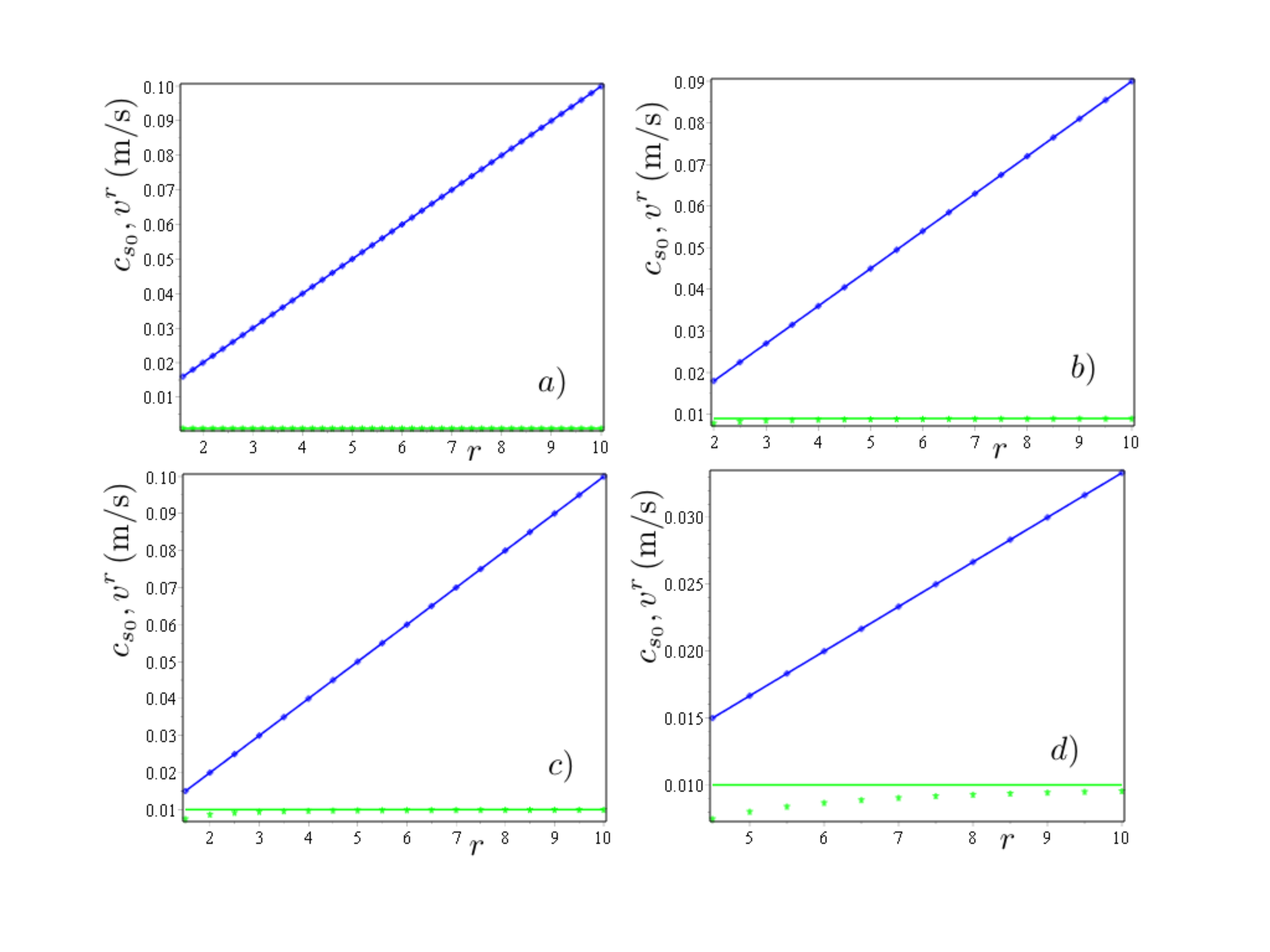}
 \caption{\footnotesize Plots for the numerical solutions $c_{s_0}(r)$ (blue diamonds) and $v^r(r)$ (green asterisks) of the system \eqref{ec:system1} and \eqref{ec:system2} and for the zero order solutions \eqref{ec:zerosolvr} for $v^r$  (green line) and \eqref{ec:zerosolcs} for $c_{s_0}$ (blue line), for several values of $v_\infty$ (in $\unit{m/s}$) and $b_0$: (a) $(0.001,0.1)$, (b) $(0.009,1)$, (c) $(0.01,1)$, (d) $(0.01,3)$. The units of $r$ are the same of the units of $b_0$. Typically, for BECs, these units are $\unit{\mu m}$.}
     \label{fig:csvrplots}
\end{figure*}
We have obtained two independent equations from $v^r$ and $c_{s_0}$, and therefore we get trivially, for \eqref{ec:system1.1} and \eqref{ec:system2.1}, respectively, the following zero order solution in $v^r/c$ and $c_{s_0}/c$
\begin{align}
\label{ec:zerosolvr}
v^r&=v_\infty,\\\label{ec:zerosolcs}
c_{s_0}&=\frac{v_\infty}{b_0}r,
\end{align}
that is, we obtain that $v^r$ is constant, and $c_{s_0}$ is linear in $r$, with slope given by $v_\infty/b_0$. 

\begin{table}[t]
  \begin{minipage}[b]{0.49\linewidth}
\centering
\begin{tabular}{c @{\hspace{15pt}} c  @{\hspace{15pt}} c}
\hline\hline
Atom & $\xi_{b,c,d}(\unit{\mu m})$ & $\xi_a(\unit{\mu m})$\\\hline
   & & \\
Li & $0.648$ & $0.324$\\
Na & $0.195$ & $0.098$\\
K   &  $0.115$ & $0.058$\\
Rb &  $0.053$ & $0.026$\\
Cs  &  $0.034$ & $0.017$\\
\hline\hline\\
\end{tabular}
\end{minipage}
  \begin{minipage}[b]{0.49\linewidth}
\centering
\begin{tabular}{c @{\hspace{15pt}} c}
\hline\hline
Plot & Spatial step ($\mu$m)\\\hline
   & \\
a)& $1.5$\\
b) & $0.6$\\
c)  &  $0.4$\\
d) &  $1.2$\\
\hline\hline\\
\end{tabular}
\end{minipage}
\caption{\footnotesize{Left: Healing length values for several alkali atoms corresponding to the velocities $c_{{s_0}_{b,c,d}}=\unit[0.01]{m/s}$ for the start of the plots b), c) and d), and $c_{{s_0}_a}=\unit[0.02]{m/s}$ for the start of the plot a). All these plots corresponding to Fig.  \ref{fig:csvrplots}. Right: Spatial step used for the plots of Fig. \ref{fig:csvrplots}.}}
\label{table:healinglength}
\end{table}
Now, we solve numerically the system without approximations, i.e. Eqs. \eqref{ec:system1} and \eqref{ec:system2}, in order to check how it fits the zero order solution.
These results are shown in Fig. \ref{fig:csvrplots}. In the lab, the values for the sound speed are typically of the order of $\unit[10^{-2}-10^{-3}]{m/s}$, therefore, in order to achieve the simulation in the lab, we only show in Fig. \ref{fig:csvrplots} plots with $c_{s_0}$ of this order.

By inspection of these plots, we found a great agreement between the numerical solutions for the complete system, i.e. without approximations, and the zero order solutions. Perhaps we found a small discrepancy for $v^r$ near the throat of the wormhole, but such discrepancy is smooth, and does not occur for all the plots, only takes places in the plot (d).

The spatial step which has been used for our plots (given in the Table 	\ref{table:healinglength}) is consistent with the healing length of the BEC, which provides the length scale above which the Gross-Pitaevskii equation and the Bogoliubov theory of the BEC work, and it is given by \cite{healinglength}
\begin{align}
\label{ec:hl}
\xi=\frac1{8\pi a\rho}=\frac{\hbar}{\sqrt{2}mc_s}
\end{align}
where we have used the relation between the scattering length $a$ and $c_s$ given by \eqref{ec:csa}. Note that here we have approximate $c_s=c_{s_0}$, due to, by \eqref{ec:cseff}, $v_\infty\ll 1$ and therefore $\gamma_s\simeq 1$. Thereby, we need a spatial resolution which is significantly larger than $\xi$.

\begin{figure*}[t!]
\includegraphics[width=0.85\textwidth]{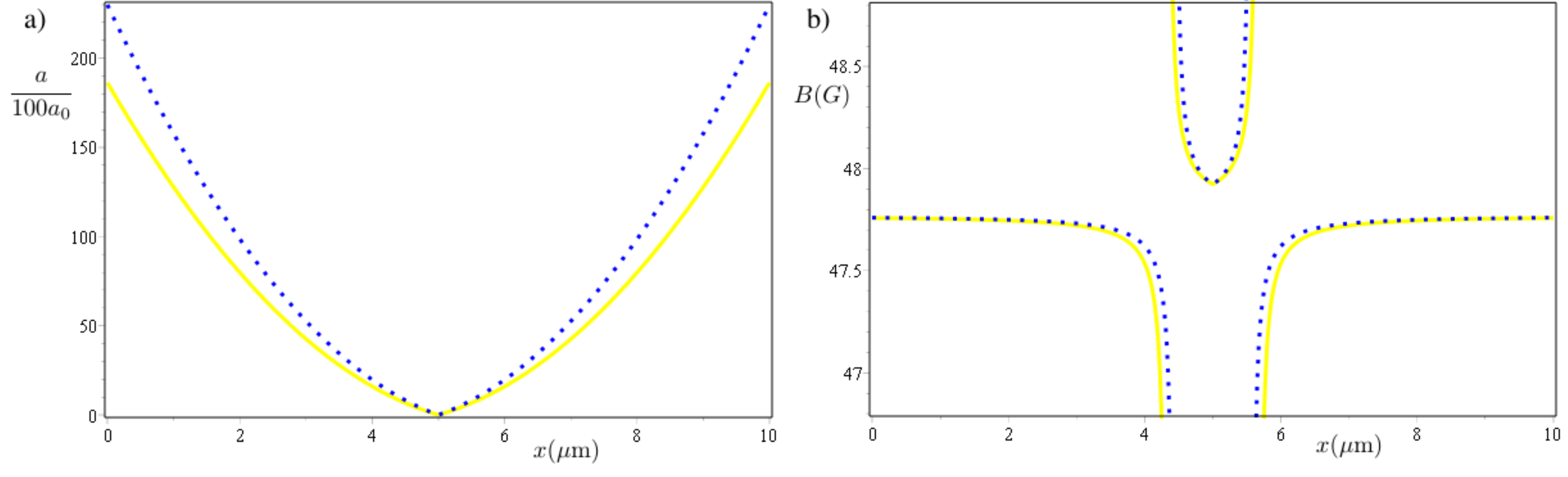}
  \caption{\footnotesize Plots for expressions (\ref{ec:ax}) and (\ref{ec:Bx}) for condensates of Cesium and for several values of $v_\infty$ and $b_0$. For each plot: $v_\infty=\unit[0.009]{m/s}$ and $b_0=\unit[1]{\mu m}$ (yellow, solid), $v_\infty=\unit[0.01]{m/s}$ and $b_0=\unit[1]{\mu m}$ (blue, dotted). We choose for these plots $R=\unit[5]{(\mu m)}$ for the size of the wormhole's branches. }
     \label{fig:aB3dplots}
\end{figure*}
Table \ref{table:healinglength} shows the values for this healing length corresponding to the initial values of $c_{s_0}$ in Fig. \ref{fig:csvrplots} for several alkali atoms which have been usually used in the context of Feshbach resonances in the lab \cite{alkali atoms}, together with the chosen spatial resolution of these plots. Note that the value of $\xi$ for the end of the plots is really irrelevant, since far from the throat the behavior of the velocities is the desired one. Therefore, due to the fact that we are only interested in the magnitude order, we only calculate $\xi$ corresponding to the initial values of the plots. Hence, as we can see in the table, the spatial step corresponding to b) and c) is one order of magnitude greater than $\xi_{b,c,d}$ for Cs and Rb, for d), the spatial step is one order of magnitude greater than this value $\xi_{b,c,d}$ for Li, Na and K, and finally the spatial step corresponding to a) is one order of magnitude greater than $\xi_a$ for Li. However, for the latter case, we have $b_0=\unit[0.1]{\mu m}$, i.e. the size of the wormhole throat is smaller than the healing length $\xi_a$, and thereby the throat would be in a more microscopic level in the BEC than the one we can consider according with the Bogoliubov theory.

In conclusion, we can simulate a wormhole spacetime seen by an observer which falls towards the hole, i.e. in the GP-like coordinates given by \eqref{ec:tr}, in condensates of Rubidium and Cesium for the profiles of the velocities of the BEC $v^r$ and $c_s$ given in \eqref{ec:zerosolvr} and \eqref{ec:zerosolcs} for the values $(v_\infty,b_0)=(0.009,1),(0.01,1)$, and in condensates of Lithium, Sodium and Potassium for the value $(v_\infty,b_0)=(0.01,3)$.

Then we want to translate the dependence in $r$ for $c_{s_0}$ that we have obtained in a radial dependence of the magnetic field $B$ and the scattering length $a$ of the BEC, which are experimentally controllable magnitudes. For this purpose, we proceed in a total analogous way to what was done in the one-dimensional case, i.e we compare our expression \eqref{ec:zerosolcs} for $c_{s_0}$ with the dependence on $B$ of the $c_s$ of a weakly interacting BEC, given in \eqref{ec:cB}. First, it is important to remark that now we have to use $c_s$ instead of $c_{s_0}$, where both are related by \eqref{ec:csgamma}, and taking into account the expression \eqref{ec:zerosolcs} for $c_{s_0}$, we have for $c_s$
\begin{align}
\label{ec:csr}
c_s=\frac{v_\infty}{b_0}r\sqrt{1-\frac{b_0^2}{r^2}}=v_\infty\sqrt{\frac{r^2}{b_0^2}-1}.
\end{align}

Now, if we equate the expressions \eqref{ec:cB} and \eqref{ec:csr} we obtain for $B(r)$
\begin{align}
\nonumber
&\tilde c_s^2\left(1-\frac{\omega}{B-B_0}\right)=v^2_\infty\left(\frac{r^2}{b_0^2}-1\right)\\\label{ec:B3d}
\Rightarrow B(r)=&\frac{\tilde c_s^2\omega}{\tilde c_s^2-v_\infty^2\left(\frac{r^2}{b_0^2}-1\right)}+B_0=\frac{\omega}{1-\frac{v_\infty^2}{\tilde c_s^2}\left(\frac{r^2}{b_0^2}-1\right)}+B_0,
\end{align}
where $\tilde c_s=(\hbar/m)\sqrt{4\pi\rho a_{bg}}$. Note that, in order to avoid confusions, we rename $c_{s0}$ in \eqref{ec:cB} as $\tilde{c}_s$.

For the Feshbach resonance, from this equation of $B(r)$ we can obtain an expression for the scattering length in terms of the spatial coordinate. Replacing \eqref{ec:B3d} in \eqref{ec:aB}
we get
\begin{align}
\label{ec:a3d}
\frac{a(r)}{a_{bg}}=\frac{v^2_\infty}{\tilde c_s^2}\left(\frac{r^2}{b_0^2}-1\right).
\end{align}

At this point, it is important to note that the radial coordinate $r$ goes from $\infty$ to $b_0$ in the upper branch of the wormhole, while in the lower branch $r$ would go from $b_0$ to $\infty$. Therefore, $r$ itself would not be the laboratory coordinate if we want to simulate both branches in the same BEC. So in order to achieve the wormhole in the laboratory, with its two branches, we need to define a different radial coordinate, in an analogous way as what we did in the one-dimensional case [see Eq. \eqref{ec:x}]. Note that at least a finite region of each branch could be accommodated in a single BEC, hence the range of the new coordinate can be finite.

Let be $x$ the new radial coordinate of the lab. We can define this coordinate as follows,
\begin{align}
\label{ec:xlab}
|x-R|=r-b_0,
\end{align}
so that the throat of the wormhole is in $x=R$. Due to the fact that $x$ must be positive, $r$ cannot be greater than $R+b_0$, and thus we have that $x$ goes from $0$ to $2R$. Thereby one branch goes from $0$ to $R$, and the other one goes from $R$ to $2R$.

Now, we can rewrite in terms of $x$ the expressions \eqref{ec:zerosolcs}, \eqref{ec:B3d} and \eqref{ec:a3d} of the magnitudes that we need for the simulation (note that $v^r$ is constant, so $v^r(r)=v^r(x)$), and we obtain respectively
\begin{align}
\label{ec:csx}
c_{s_0}(x)&=\frac{v_\infty}{b_0}\left(|x-R|+b_0\right),\\
\label{ec:Bx}
B(x)&=\frac{\omega}{1-\frac{v_\infty^2}{\tilde c_s^2}\left(\frac{(|x-R|+b_0)^2}{b_0^2}-1\right)}+B_0,\\
\label{ec:ax}
\frac{a(x)}{a_{bg}}&=\frac{v^2_\infty}{\tilde c_s^2}\left(\frac{(|x-R|+b_0)^2}{b_0^2}-1\right).
\end{align}

Finally, we plot the expressions for the scattering length and the magnetic fields in terms of $x$, given by \eqref{ec:ax} and \eqref{ec:Bx}, respectively, in order to compare with the experimental spatial variation of $a$, given in \cite{Exp}, as in the one-dimensional case. We particularise these expressions for Cesium condensates, which is the element used in \cite{Exp}, and we take again the same parameters $a_{bg}$, $\omega$ and $B_0$ as before. For the density of the BEC, we use the typical value $\rho=\unit[10^{15}]{cm^{-3}}$\cite{Barcelo, dalfovo}. On the other hand, for the parameters relative to the wormhole $v_\infty$ and $b_0$, we use the values of these quantities from plots \ref{fig:csvrplots} which are in agreement with the healing length for Cs, i.e. $(v_\infty, b_0)=(0.009,1), (0.01, 1)$.

    In Fig. \ref{fig:aB3dplots} (a) we see that the behavior of the spatial dependence of $a$ is similar as the one-dimensional case, but now this quantity reaches higher values than in the experimental plots. This fact could imply that our wormhole is not realizable in BECs of Cs with current technology. Moreover, two asymptotes arise in Fig.  \ref{fig:aB3dplots} (b), rendering a magnetic field profile which seems experimentally challenging. It seems likely that considering more general values of the parameter $q$ might result in more amenable profiles for the external magnetic field and the scattering length, as in the one-dimensional case. We leave the exploration of this idea for future research.

\section{Conclusions}

We provide a recipe to perform a quantum simulation of  wormhole spacetimes in weakly interacting BECs, both in $1+1$ and $3+1$ dimensions.

In the one-dimensional case, we propose a profile for the external magnetic field and the scattering length of the BEC in terms of the spatial coordinate, which allows to simulate a family of wormhole spacetimes corresponding to the range $0<q<1$.  We show that this is within reach of current state-of-the-art technologies.

On the other hand, in the three-dimensional case we present a solution which enables to build up a quantum simulator of an Ellis wormhole spacetime in generalized Gullstrand-Painlev\'e coordinates. We show a simple and elegant form for the speed of the phonons of the condensate, and a corresponding expression for the magnetic field and the scattering length, from which the simulation can be achieved. However, the experimental requirements in this case seem to go beyond current capabilities.

\section*{Acknowledgements}
Financial support from Fundaci\'on General CSIC (Programa ComFuturo) is acknowledged by C.S.

\end{document}